\documentclass[12pt]{iopart}
\usepackage{iopams}
\usepackage{bm}

\begin{document}

\newcommand{\beq}{\begin{equation}}
\newcommand{\eeq}{\end{equation}}
\newcommand{\barr}{\begin{eqnarray}}
\newcommand{\earr}{\end{eqnarray}}

\newcommand{\andy}[1]{ }
\def\cH{{\cal H}}
\def\cV{{\cal V}}
\def\cU{{\cal U}}
\def\bra#1{\langle #1 |}
\def\ket#1{| #1 \rangle}

\def\cH{{\cal H}}
\def\coltwovector#1#2{\left({#1\atop#2}\right)}
\def\upp{\coltwovector10}
\def\downn{\coltwovector01}
\def\Ord{\mbox{O}}
\def\bmp{\mbox{\boldmath $p$}}
\def\rhobar{\bar{\rho}}
\renewcommand{\Re}{{\rm Re}}
\renewcommand{\Im}{{\rm Im}}

\def\ask{\marginpar{?? ask:  \hfill}}
\def\fin{\marginpar{fill in ... \hfill}}
\def\note{\marginpar{note \hfill}}
\def\check{\marginpar{check \hfill}}
\def\discuss{\marginpar{discuss \hfill}}

\title[Zeno dynamics and constraints]{Zeno dynamics and constraints}

\author{P.\ Facchi\dag
\footnote[3]{To
whom correspondence should be addressed
(paolo.facchi@ba.infn.it)}, G.\ Marmo\ddag, S. Pascazio\dag, A.\
Scardicchio\P\ and E.C.G. Sudarshan$\sharp$ }

\address{\dag\ Dipartimento di Fisica, Universit\`a di Bari \\
       and Istituto Nazionale di Fisica Nucleare, Sezione di Bari,
 I-70126  Bari, Italy}

\address{\ddag\ Dipartimento di Scienze Fisiche, Universit\`a Federico II di Napoli
        \\ and Istituto Nazionale di Fisica Nucleare, Sezione di
       Napoli, I- 80126  Napoli, Italy}

\address{\P\ Center for Theoretical Physics, Massachusetts Institute of
Technology, Cambridge, MA 02139, USA}

\address{$\sharp$\ Physics Department, University of Texas at Austin
 Texas 78712, USA}

\begin{abstract}
We investigate some examples of quantum Zeno dynamics, when a
system undergoes very frequent (projective) measurements that
ascertain whether it is within a given spatial region. In
agreement with previously obtained results, the evolution is found
to be unitary and the generator of the Zeno dynamics is the
Hamiltonian with hard-wall (Dirichlet) boundary conditions. By
using a new approach to this problem, this result is found to be
valid in an arbitrary $N$-dimensional compact domain. We then
propose some preliminary ideas concerning the algebra of
observables in the projected region and finally look at the case
of a projection onto a lower dimensional space: in such a
situation the Zeno ansatz turns out to be a procedure to impose
constraints.
\end{abstract}



\maketitle

\section{Introduction}

Very frequent measurement can slow the time evolution of quantum
mechanical systems. This is, in a few words, the quantum Zeno
effect (QZE), by which transitions to states different from the
initial one are gradually suppressed as the measurement frequency
$N$ becomes very large \cite{QZEseminal,MS} (for a review, see
\cite{QZEreview}). There are, however, two important issues that
deserve attention: firstly, for a general (incomplete and
nonselective \cite{Schwinger}) measurement, represented by a
complete set of projections onto multidimensional subspaces
(rather than a single-dimensional one, as in the usual formulation
of the QZE, by which the measurement ascertains whether the system
is still in its initial, pure state), the quantum system may---and
indeed does---evolve away from its initial state, although it
remains in the subspace defined by the measurement (and
represented by a multidimensional projection operator)
\cite{compact,regularize}. This leads to the formation of the
``Zeno subspaces" \cite{FPsuper}. Secondly, if the measurement is
not \emph{very} frequent, the quantum evolution yields the
so-called ``inverse" or ``anti" Zeno effect, by which transitions
away from the initial state (or in general out of the relevant
subspaces) are accelerated \cite{antiZeno}.

Both the Zeno and inverse Zeno phenomena have been experimentally
observed during the last few years
\cite{WinelandZeno,Wilkinson,raizenlatest,Balzer} (but
see \cite{valanju} for previous analyses of experimental data on
nuclear hadronic cascades). The first experiment was done with an
oscillating system
\cite{WinelandZeno}, according to an interesting proposal by Cook
\cite{Cook}, and was widely debated
\cite{Itanodisc}. In a recent beautiful set of experiments,
performed by Raizen's group, first the initial quadratic and
non-Markovian Zeno region was observed \cite{Wilkinson}, then both
the quantum Zeno and inverse Zeno effects were proved for
\emph{bona fide} unstable system (probability leakage out of an
optical potential) \cite{raizenlatest}.

In this article we shall mainly analyze the first issue,
investigating the features of the Zeno (sub)dynamics in the
relevant subspace. This and related problems were contemplated in
the seminal formulation of the QZE \cite{MS}, where it was proved
that the dynamics is governed by a semigroup. The details of the
dynamics had interesting and challenging mathematical aspects,
that were independently investigated by other authors
\cite{Friedman72,Gustavson}. As a matter of fact, some
mathematical issues are still unresolved nowadays. One of the most
intruiguing features of the original paper \cite{MS} is that some
delicate operator properties were postulated on physical grounds;
curiously, these postulates are always found to be valid in
concrete examples, even nontrivial ones.

For a wide class of measurements, namely those represented by
spatial projections, one can prove that the system evolves
unitarily in a proper subspace of the total Hilbert space, the
generator of the dynamics being the Hamiltonian with Dirichlet
boundary conditions on the region associated with the spatial
projection
\cite{compact,regularize}. This finding motivated further
interesting studies on this topic
\cite{MisraAntoniou,GustavsonSolvay,Schmidt,ExnerIchinose}. In
particular, Exner and Ichinose \cite{ExnerIchinose} analyzed this
result in a rigorous framework, under the nontrivial (and
interesting) assumption that the original Hamiltonian be lower
bounded and the Zeno Hamiltonian densely defined in the Hilbert
space. The aim of this article is to further elaborate on these
issues. We will first explicitly work out some
examples---essentially the free case in two and three dimensions,
with projections onto regular domains---and introduce a novel
calculation technique, giving a constructive proof of the Zeno
Hamiltonian. We then extend this result to a general spatial
projection in $N$ dimensions.

We shall prove that the Dirichlet boundary conditions are a
consequence of the Zeno procedure (different proofs can be given,
at different levels of generality and mathematical rigor, see
\cite{Friedman72,compact,regularize,ExnerIchinose}), by
exploring an interesting method of calculation,
based on asymptotic techniques, that yields a stationary
Schr\"odinger equation with the appropriate (Dirichlet) boundary
conditions for its eigenfunctions. In Sec.\ \ref{sec:gener} we set
up the general framework and introduce notation. In Sec.\
\ref{sec:strip} the projection domain is a rectangle in the plane.
In Sec.\ \ref{sec:annulus} it is an annulus in the plane. In Sec.\
\ref{sec:sphere} we look at a spherical shell in $\mathbb{R}^3$.
In Sec.\ \ref{sec:generalization} we generalize to regular domains
in $\mathbb{R}^N$ and in Sec.\ \ref{sec:algebra} we briefly
discuss the Zeno dynamics in the Heisenberg picture as well as the
features of the algebra of observables in the projected domain. In
Sec.\ \ref{sec:tuttoins} we look at a different case, when the
system is projected onto a domain of lower dimensionality: we
shall only look at some examples and shall not attempt to
generalize. One can say that in this case the Zeno ansatz yields a
procedure to impose a constraint. The ideas we propose in these
last two sections are somewhat embryonic and can be considered as
plans for future developments. In Sec.\
\ref{sec:concl} we comment on future perspectives and
applications.

\section{Zeno subdynamics}
\andy{gener}
\label{sec:gener}
Consider a free particle in $N$-dimensions with the Hamiltonian
\beq
\label{eq:HU}
H=\frac{\bm p^2}{2 M}=-\frac{\hbar^2\triangle}{2 M}, \qquad
U(t)=e^{-iHt/\hbar}
\eeq
acting on $\psi \in L^2(\mathbb{R}^N)$. Given a compact domain
$D\subset\mathbb{R}^N$ with a nonempty interior and a regular
boundary, consider the projection operator
\beq
\label{eq:P}
P=\chi_D(\bm x)=\int_D d^N \bm x \ket{\bm x}\bra{\bm x}, \qquad P
\psi(\bm x) = \chi_D(\bm x) \psi(\bm x),
\eeq
where $\chi_D(\bm x)$ is the characteristic function of the domain
$D$, and thought of as an operator, along with its complement
$Q=1-P=1-\chi_D(\bm x)$, decomposes the space $L^2(\mathbb{R}^N)$
into orthogonal subspaces. The Zeno subdynamics evolution operator
is given by the limit
\beq
\label{eq:UZ}
U_Z(t) = \lim_{N\to\infty}\left(G(t/N)\right)^N,
\eeq
where the (nonunitary) evolution
\beq
\label{eq:1step}
G(\tau)=P U(\tau) P
\eeq
represents a single step (projection-evolution-projection) Zeno
process.

Under rather general hypotheses the limit (\ref{eq:UZ}) can be
proved to exist
\cite{Friedman72,MS,compact,regularize,ExnerIchinose} and yields a
unitary evolution group in a proper subspace of $L^2(D)$. One gets
\beq
U_Z(t)= P \exp(-i H_Z t/\hbar),
\label{eq:UZ1}
\eeq
where the generator of the dynamics is the Zeno Hamiltonian
\beq
\label{eq:HZ}
H_Z =-\frac{\hbar^2\triangle_D}{2 M},
\eeq
defined in the domain
\beq
\label{eq:domH}
D(H_Z)=\{\psi\in L^2(D)\;|\;
\triangle\psi\in L^2(D), \psi(\partial D)=0\},
\eeq
$\partial D$ being the boundary of $D$ (hard-wall or Dirichlet
boundary conditions).

We will focus on this problem by looking for the eigenbasis
$\{\ket{{\bm n}}\}$ of $U_Z(t)$ in the subspace
$PL^2(\mathbb{R}^N)\simeq L^2(D)$  such that
\beq
\label{eq:weaklimit}
\bra{{\bm n}}U_Z(t)\ket{{\bm m}}=\lim_{N\to\infty} \bra{{\bm
n}}G(t/N)^N\ket{{\bm m}}=\bra{{\bm n}}e^{-iH_Z t/\hbar}\ket{{\bm
m}}=\delta_{\bm m ,\bm n}e^{-iE_{\bm n} t/\hbar}.
\eeq
In order to find this basis consider an arbitrary orthonormal
complete set of functions in $L^2(D)$
\beq
\Psi_{\bm n}(\bm x)=\bra{\bm x} \bm n\rangle
\eeq
and take the matrix elements of the single-step operator
(\ref{eq:1step})
\beq
\label{eq:matrixel}
G_{\bm m,\bm n}(t)=\bra{\bm m}G(t)\ket{\bm n} =
\Tr \left[G(t)\ket{\bm n}\bra{\bm m} \right].
\eeq
If the matrix elements of the single-step operator behave like
\beq
\label{eq:diag}
G_{\bm m ,\bm n}(t) =\delta_{\bm m, \bm n}\left(1-i\frac{E_{\bm
n}t}{\hbar}\right)+ R_{\bm m ,\bm n}(t),
\eeq
where for $t\to0$
\beq
\label{eq:diagbis}
R_{\bm m,\bm n}(t)=o(t),
\eeq
then, under the assumption of uniform convergence of the infinite
sums stemming from the insertion of $N-1$ resolutions of the
identity in (\ref{eq:UZ}), one obtains:
\barr
G_{\bm m,\bm n}^Z(t)&\equiv&\bra{\bm m}U_Z(t)\ket{\bm n}
\nonumber\\
&=&\lim_{N \to \infty} \sum_{\bm n_1,\ldots, \bm n_{N-1}} G_{\bm
m, \bm n_1}(t/N)G_{\bm n_1,\bm n_2}(t/N) \cdots G_{\bm n_{N-1},\bm
n}(t/N)
\nonumber\\
&=& \delta_{\bm m, \bm n} \exp\left(-i\frac{ E_{\bm n}
t}{\hbar}\right).
\earr
The basis $\{\ket{{\bm n}}\}$ is thus the eigenbasis of $H_Z$
belonging to the eigenvalues $E_{\bm n}$:
\beq
\label{eq:eigenequ}
H_Z \Psi_{\bm n}(\bm x) = E_{\bm n} \Psi_{\bm n}(\bm x).
\eeq
Notice that when we apply $U(t/N)$ to the relevant subspace
$PL^2(\mathbb{R}^N)$, the transformed space need not be orthogonal
anymore to $Q L^2(\mathbb{R}^N)$, where $Q=1-P$, and the
$t/N$-dependence of the scalar product of two vectors in these two
subspaces is given by
\beq
Q U(t/N)P= O(t/N).
\label{eq:nonorth}
\eeq
It has been shown that Eq.\ (\ref{eq:diag}) implies Dirichlet
boundary conditions for the states $\Psi_{\bm n}(\bm x)$ (the
``Zeno eigenbasis") \cite{regularize}. The proof, based on
asymptotic techniques, yields the propagator in an appropriately
chosen basis of eigenfunctions. In the following sections we shall
introduce a novel approach: by using asymptotic analysis and the
path integral representation of the matrix element
(\ref{eq:matrixel}), we will obtain a stationary Schr\"odinger
equation and a set of boundary conditions for its eigenfunctions.
This will enable us to define the induced Zeno Hamiltonian $H_Z$
and its spectrum. The advantage of the present approach, as
compared to the previous one \cite{regularize}, lies in the fact
that one can derive a Schr\"odinger equation with Dirichlet
boundary conditions in the projected (Zeno) subspace. Moreover, by
examining some examples of multiple connected domains, we will
show how the Zeno dynamics induces constraints that inherit the
topological properties of the parent space.

\section{Rectangle}
\andy{strip}
\label{sec:strip}

We start off with one of the simplest examples and introduce the
procedure. Consider a rectangle in the plane, $D=[0,a]\times
[0,b]\subset\mathbb{R}^2$. In this case the projection
(\ref{eq:P}) reads
\beq
P=\chi_{[0,a]}(x)\; \chi_{[0,b]}(y)= \int_0^a dx \int_0^b dy
\ket{xy}\bra{xy}
\eeq
and the Hamiltonian (\ref{eq:HU}) is
\barr
H=\frac{p_x^2+p_y^2}{2M}=-\frac{\hbar^2}{2M}(\partial_x^2+\partial_y^2).
\earr
The Zeno Hamiltonian, engendering the Zeno subdynamics, is
formally given by (\ref{eq:HZ})-(\ref{eq:domH}) and represents a
free particle in the box $D=[0,a]\times [0,b]$ with Dirichlet
boundary conditions
\barr
\label{eq:HZstrip}
& & H_Z=-\frac{\hbar^2}{2M}(\partial_x^2+\partial_y^2), \\
& & \qquad
\psi(0,y)=\psi(a,y)=0,\quad \psi(x,0)=\psi(x,b)=0.
\label{eq:HZstripb}
\earr
The eigenfunctions and eigenvalues are well known
\barr
\label{eq:eigenstrip}
& & \Psi_{n
m}(x,y)=\sqrt{\frac{2}{a}}\sin\left(\frac{n\pi}{a}x\right)
\sqrt{\frac{2}{b}}\sin\left(\frac{n\pi}{b}y\right), \\
& &  E_{n m}=\frac{\hbar^2
\pi^2}{2M}\left(\frac{n^2}{a^2}+\frac{m^2}{b^2}\right).
\label{eq:eigenstripb}
\earr
Let us look in detail at the derivation of the Zeno subdynamics
(\ref{eq:HZstrip})-(\ref{eq:eigenstripb}) in this particular case.
As explained in Sec.\
\ref{sec:gener}, the eigenbasis of the Zeno Hamiltonian $H_Z$ in
$L^2(D)$,
\beq
\Psi_{nm}(x,y)=\bra{x,y} nm\rangle
\eeq
must satisfy condition (\ref{eq:diag}):
\beq
\label{eq:diagstrip}
G_{n'm',nm}(t)=\delta_{n'n}\delta_{m'm}\left(1-i\frac{E_{nm}t}{\hbar}\right)+
o(t),
\eeq
where
\beq
G_{n'm',nm}(t)=\bra{n'm'}G(t)\ket{nm}.
\eeq
are the matrix elements (\ref{eq:matrixel}) of the single-step
evolution operator.

This can be proved by direct inspection: one gets
\barr
G_{n'm',nm}(t)&=&\int_0^a dx\int_0^b dy \int_0^a dx'\int_0^b
dy'\left(\frac{M}{2\pi i\hbar
t}\right)e^{i\frac{M((x'-x)^2+(y'-y)^2)}{2\hbar t}} \nonumber \\
& & \times
\Psi^*_{n'm'}(x,y)\Psi_{nm}(x',y')
\label{eq:rettangolo}
\earr
and by substituting $\xi=x'-x$ and $\eta=y'-y$
\barr
G_{n'm',nm}(t)&=&\int_0^a dx\int_0^b dy\int_{-x}^{a-x} d\xi
\int_{-y}^{b-y} d\eta \left(\frac{M}{2\pi i\hbar
t}\right)e^{i\frac{M(\xi^2+\eta^2)}{2\hbar t}} \nonumber \\
& & \times
\Psi^*_{n'm'}(x,y)\Psi_{nm}(x+\xi,y+\eta).
\earr
With the natural choice $\Psi_{n m}(x,y)=\psi_n(x)\phi_m(y)$ this
yields the product of two quantities
\barr
G_{n'm',nm}(t)&=& G_{n'n}(t) G_{m' m}(t)
\nonumber \\
& = & \int_0^a dx \int_{-x}^{a-x} d\xi
\left(\frac{M}{2\pi i\hbar t}\right)^{1/2}
e^{i\frac{M\xi^2}{2\hbar t}}
\psi^*_{n'}(x)\psi_{n}(x+\xi)\nonumber\\
& &\times\int_0^b dy' \int_{-y}^{b-y} d\eta \left(\frac{M}{2\pi
i\hbar t}\right)^{1/2}e^{i\frac{M\eta^2}{2\hbar
t}}\phi^*_{m'}(y)\phi_{m}(y+\eta) \;\;
\earr
and accordingly $E_{nm}=E_n+E_m$. Consider the first quantity
$G_{n'n}$ and the integral over $\xi$. In the small-$t$ limit
there are contributions from the boundary points $\xi=-x$ and
$\xi=a-x$ and from the stationary point $\xi=0$
\beq
G_{n'n}= \int_0^a dx \psi^*_{n'}(x) [\mbox{bound}+\mbox{stat}],
\eeq
where
\barr
\mbox{bound}&=&\left(\frac{M}{2\pi i\hbar
t}\right)^{1/2}\frac{\hbar t }{i M
\xi}\psi_n(x+\xi)e^{i\frac{M\xi^2}{2\hbar
t}}\Bigg|_{\xi=-x}^{\xi=a-x}+O\left(t^{3/2}\right)\nonumber\\
&&=\sqrt{\frac{\hbar t}{-2\pi iM}}\left(\frac{e^{iM(x-a)^2/2\hbar
t}}{x-a}\psi_n(a)-\frac{e^{iMx^2/2\hbar
t}}{-x}\psi_n(0)\right)+O\left(t^{3/2}\right), \nonumber \\
\label{eq:denomworries}
\earr
while ($\lambda=M/2\hbar t$)
\barr
\mbox{stat}&=&\int_{-\infty}^\infty d\xi\left(\psi_n(x)+
\psi_n'(x)\xi+\frac{1}{2!}\psi_n''(x)\xi^2+O(\xi^3)\right)\sqrt{\frac{\lambda}{\pi
i}} e^{i \lambda \xi^2}\nonumber\\
&=&\psi_n(x)+i\frac{t\hbar}{2M}\psi_n''(x)+O(t^{2}).
\earr
In order to obtain (\ref{eq:diagstrip}) one must require that
(remember that $E_{nm}=E_n+E_m$)
\barr
\mathrm{bound}= O(t^{3/2})\qquad \mathrm{and}\quad
-\frac{\hbar}{2M}\psi_n''(x)&=&\frac{E_n}{\hbar}\psi_n(x),
\earr
which translates into
\barr
\label{eq:diffbc}
-\frac{\hbar^2 }{2M}\partial_x^2\psi_n(x)=E_n\psi_n(x), \qquad
\mathrm{with}\quad \psi_n(0)=\psi_n(a)=0,
\earr
so that for $G_{n'n}$ one obtains
\beq
G_{n'n}(t)=\left(1-i\frac{E_n t}{\hbar}\right)\delta_{n'n}+
O(t^{3/2})
\eeq
and analogously for $G_{m'm}$, so that
\beq
G_{n'm',nm}(t)=\left(1-i\frac{E_n t}{\hbar}-i\frac{E_m
t}{\hbar}\right)\delta_{n'n}\delta_{m'm}+ O(t^{3/2}),
\eeq
which has exactly the form (\ref{eq:diagstrip}). By Eq.\
(\ref{eq:diffbc}) and its analog for $\phi_m(y)$, the
eigenfunctions $\Psi_{nm}(x,y)=\psi_n(x)\phi_m(y)$ of $H_Z$
satisfy
\barr
-\frac{\hbar^2
}{2M}(\partial_x^2+\partial_y^2)\Psi_{n,m}(x,y)=(E_n+E_m)\Psi_{n,m}(x,y),
\earr
with Dirichlet boundary conditions. They are therefore given by
(\ref{eq:eigenstrip}). The Zeno Hamiltonian is therefore
(\ref{eq:HZstrip})-(\ref{eq:HZstripb}).

This derivation, although it yields the desired (and correct)
result, is not rigorous. In particular, it does not tackle the
delicate problem of understanding the convergence properties of
the asymptotic expansion at the intersection of the $(x,y)$ and
$(x',y')$ boundaries in Eq.\ (\ref{eq:rettangolo}) [this is
apparent if one looks at the denominators of the far right hand
side of Eq.\ (\ref{eq:denomworries})]. A similar approach will be
adopted in the next sections. A more rigorous proof can be given,
but will not be presented here.

\section{Annulus}
\andy{annulus}
\label{sec:annulus}

Consider now a circular annulus (or ring) of width $\delta
r=r_2-r_1$ on the plane, defining the domain
$D=\{(x,y)\;|\;r_1^2\leq x^2+y^2
\leq r_2^2\}$. The projection on $D$ reads
\beq
P=\chi_{[r_1,r_2]}(r)=\int_D dx dy
\ket{xy}\bra{xy}=\int_{r_1}^{r_2} dr r \int_0^{2\pi} d\theta
\ket{r \theta}\bra{r\theta}.
\eeq
\beq
r_2-r_1\equiv\delta r>0.
\eeq

The Zeno Hamiltonian, engendering the Zeno subdynamics, is given
by (\ref{eq:HZ}) and represents a free particle on $D$ with
Dirichlet boundary condition
\barr
& & H_Z=-\frac{\hbar^2}{2M}(\partial_x^2+\partial_y^2)
=-\frac{\hbar^2}{2M}\left(\frac{1}{r}\partial_r(r
\partial_r)+\frac{1}{r^2}\partial_\theta^2\right) ,\label{eq:HZringa} \\
& &
\qquad \psi(r_1,\theta)=\psi(r_2,\theta)=0.
\label{eq:HZring}
\earr
As is well known, by writing the eigenfunctions of $H_Z$ as
$\Psi_{nl}(r,\theta)=\psi_{nl}(r)\phi_l(\theta)$, the angular
functions are given by
\beq
\phi_l(\theta)=\frac{1}{\sqrt{2\pi}} \exp(i l \theta),
\qquad \mathrm{with}
\quad l=0,\pm1,\pm2,\ldots,
\eeq
while the radial part of the eigenvalue equation reads
\barr
\label{eq:eigenring}
& & -\frac{\hbar^2}{2M}\frac{1}{r}\partial_r(r
\partial_r)\psi_{nl}(r)+\frac{\hbar^2 l^2}{2M
r^2}\psi_{nl}(r)=E_{nl}\psi_{nl}(r), \\
& &
\label{eq:eigenringb}
\qquad \psi_{nl}(r_1)=\psi_{nl}(r_2)=0
\earr
and can be solved in terms of Bessel functions.

Let us look in detail at the derivation of the Zeno subdynamics
(\ref{eq:HZringa})-(\ref{eq:eigenringb}) in this particular case.
As explained in Sec.\ \ref{sec:gener}, the eigenbasis of the Zeno
Hamiltonian  $H_Z$ in $L^2(D)$,
\beq
\Psi_{nl}(r,\theta)=\bra{r,\theta} nl\rangle
\eeq
must satisfy condition (\ref{eq:diag}), that is
\beq
\label{eq:diagring}
G_{n'l',nl}(t)=\delta_{n'n}\delta_{l'l}\left(1-i\frac{E_{nl}t}{\hbar}\right)+
o(t),
\eeq
where
\beq
G_{n'l',nl}(t)=\bra{n'l'}G(t)\ket{nl}
\eeq
are the matrix elements (\ref{eq:matrixel}) of the single-step
evolution operator.

By writing $\Psi_{n l}(r,\theta)=\psi_{nl}(r)\phi_l(\theta)$, we
get
\barr
G_{n'l',nl}(t)&=&\int_{r_1}^{r_2}r dr\int_{r_1}^{r_2}r'
dr'\int_0^{2\pi}d\theta \int_0^{2\pi}d\theta' \nonumber \\
& &
\times \psi_{nl}(r)\phi_l(\theta)\psi^*_{n'l'}(r')\phi^*_{l'}(\theta')
\left(\frac{M}{2\pi i \hbar t}\right) e^{i\frac{M d^2}{2\hbar t}},
\earr
where $d$ is the distance between the points $(r,\theta)$ and
$(r',\theta')$
\beq
d^2=r'^2+r^2-2r'r\cos(\theta'-\theta)=(r'-r)^2+2r'r(1-\cos(\theta'-\theta)).
\eeq
Let us look first at the $\theta$ integrals. Changing again to
$\eta=\theta-\theta'$ and dropping the prime one gets
\barr
G_{n'l',nl}(t)&=&\int_{r_1}^{r_2}r dr\int_{r_1}^{r_2}r' dr'
\psi_{nl}(r)\psi^*_{n'l'}(r')\left(\frac{M}{2\pi i \hbar
t}\right)^{1/2}e^{i\frac{M(r'-r)^2}{2\hbar t}}\nonumber\\
&&\times \int_0^{2\pi}d\theta
\int_{-\theta}^{2\pi-\theta}d\eta\;\phi_l(\theta+\eta)\phi^*_{l'}(\theta)
\left(\frac{M}{2\pi i \hbar t}\right)^{1/2} e^{i\frac{Mr'r}{\hbar
t}(1-\cos\eta)}. \nonumber \\
\earr
Consider the integral over $\eta$ (at fixed $r'$ and $r$). In the
limit $t\to0$ the boundary contribution reads ($z=Mr'r/\hbar t$)
\beq
\mbox{bound}=\frac{1}{\sqrt{r'r}}\frac{-i}{\sqrt{2\pi i z
\sin^2\theta}}e^{i z
(1-\cos(\theta))}\left[\phi_l(2\pi)-\phi_l(0)\right]+O(t^{3/2}).
\eeq
In order that $O(\sqrt{t})$ vanishes and (\ref{eq:diagring}) is
satisfied, one must require the periodicity
\beq
\phi_l(0)=\phi_l(2\pi).
\eeq
The difference with the preceding case is given by the periodicity
of the Green function.

However, we now have \emph{two} stationary points in the $\eta$
integral. One is $\eta=0$ and the other is $\eta=\pi$ for
$\theta<\pi$, or $\eta=-\pi$ for $\theta>\pi$. These represent the
minimum and maximum of the distance between two points, one fixed
on the circle $r'=\mbox{const.}$ and the other one located on the
circle $r=\mbox{const.}$ at an angle $\eta$. One should get (at
least) two points of stationary phase each time one constrains the
dynamics on a closed (iper)surface. Both contributions must be
taken into account. The only difference with the previous case is
that one must consider also $\eta^4$ terms arising from the cosine
in the integral
\barr
\mbox{stat}_0&=&\frac{1}{\sqrt{r'r}}\int d\eta\sqrt{\frac{z}{2\pi
i}}\left[
\phi_l(\theta)+\frac{1}{2!}\phi''_l(\theta)\eta^2-i\frac{z}{4!}\eta^4
\phi_l(\theta)\right]e^{iz\eta^2/2} , \\
\mbox{stat}_{\pm\pi}&=&\frac{1}{\sqrt{r'r}}e^{2iz}\int
d\eta\sqrt{\frac{z}{2\pi i}}\left[
\phi_l(\theta\pm\pi)+\frac{1}{2!}\phi''_l(\theta\pm\pi)\eta^2
\right.
\nonumber \\
& & \left. \ + \; i\frac{z}{4!}\eta^4
\phi_l(\theta\pm\pi)\right]e^{-iz\eta^2/2} .
\earr
Notice that stat$_{\pm\pi}$ has a phase $2z=2mr'r/\hbar t$. This
phase changes the term $m(r'-r)^2/2 \hbar t$ of the $r',r$
integrals into a term $m(r'+r)^2/2\hbar t$. This factor has no
more stationary points in the $r',r$ integrals, so that its
contribution can be neglected (in the $t\to 0$ limit). In turn,
also the contribution from stat$_{\pm\pi}$ can be neglected. On
the other hand, the stat$_0$ contribution is
\beq
\mbox{stat}_0=\frac{1}{\sqrt{r'r}}\left[\phi_l(\theta)+\frac{i\hbar
t}{2Mr'r}\phi''_l(\theta)+\frac{i\hbar
t}{8Mr'r}\phi_l(\theta)\right]+O(t^2)
\eeq
and following the same reasoning as in Sec.\ \ref{sec:gener}
(rectangle on the plane) one obtains a differential equation for
the eigenfunctions
\barr
-\phi''_l(\theta)=\alpha_l\phi_l(\theta),
\qquad
\phi(0)=\phi(2\pi)
\earr
which yields $\alpha_l=l^2$, whence
\beq
\int_0^{2\pi}d\theta\phi^*_{l'}(\theta)\;
\mathrm{stat}_0=\delta_{l'l}\frac{1}{\sqrt{r'r}}\left[1-i t
\frac{\hbar}{2Mr'r}\left(l^2-\frac{1}{4}\right)\right]+O(t^2).
\eeq
Therefore the integral over $r',r$ reads
\barr
& & \int_{r_1}^{r_2}rdr\int_{r_1}^{r_2}r'dr'\sqrt{\frac{M}{2\pi
i\hbar t}}\psi^*_{n'l'}(r')\psi_{nl}(r)e^{i\frac{M(r'-r)^2}{2\hbar
t}}\delta_{l'l} \nonumber \\
& & \quad \times \frac{1}{\sqrt{r'r}}\left[1-i t
\frac{\hbar}{2Mr'r}\left(l^2-\frac{1}{4}\right)\right].
\earr
By inserting $\xi=r-r'$ and dropping the prime on $r'$ we get
\barr
& &
\int_{r_1}^{r_2}\sqrt{r}dr\int_{r_1-r}^{r_2-r}\sqrt{\xi+r}d\xi\sqrt{\frac{M}{2\pi
i\hbar t}}\psi^*_{n'l'}(r)\psi_{nl}(r+\xi)e^{i\frac{M\xi^2}{2\hbar
t}}\delta_{l'l} \nonumber \\
& & \quad \times \left[1-i t
\frac{\hbar}{2Mr(r+\xi)}\left(l^2-\frac{1}{4}\right)\right].
\earr
By the same reasoning as before one obtains a differential
equation and the Dirichlet boundary conditions for the functions
$A_{nl}(r)=\sqrt{r}\psi_{nl}(r)$:
\barr
\label{eq:Anla}
& & -\frac{\hbar^2}{2M}A_{nl}''(r)+\frac{\hbar^2}{2Mr^2}
\left(l^2-\frac{1}{4}\right)A_{nl}(r)=E_{nl}A_{nl}(r), \\
& & \qquad \qquad A_{nl}(r_1)=A_{nl}(r_2)=0 .
\label{eq:Anl}
\earr
In terms of  the radial functions $\psi_{nl}$, Eq.\
(\ref{eq:Anla}) becomes just Eq.\ (\ref{eq:eigenring}), whence the
Zeno Hamiltonian is given by (\ref{eq:HZringa})-(\ref{eq:HZring}).

It is interesting to notice that in this case of multiple
connectedness the Zeno dynamics yields no Aharonov-Bohm
topological phases. In a few words, one might say that the
projected dynamics on the annulus ``inherits" the topological
properties of the initial space $\mathbb{R}^2$, and in particular
the single valuedness of the wave function. The spatial
projections do not introduce any additional ``twist" into the
system, that could induce a phase.

Two additional quick comments: first, the $r_1 \to 0$ limit yields
a circle; however, it does \emph{not} yield the Zeno dynamics on
the domain $D=\{(x,y)|x^2+y^2<r_2^2 \}$, because of the spurious
condition $\psi_{nl}(0)=0$, excluding $s$-wave eigenfunctions.
This seemingly trivial remark clarifies that taking a limit of the
projected domain does \emph{not} necessarily yield the right Zeno
dynamics. Second, the circular ring sector $\{(r,\theta)|r_1\leq r
\leq r_2, \theta_1 \leq \theta \leq \theta_2 \}$ can be easily
computed and yields the right dynamics and eigenfunctions (Bessel
functions $I_\mu(r), \mu \in \mathbb{R}$) \cite{Sommerfeld}.

\section{Spherical shell}
\andy{sphere}
\label{sec:sphere}

Let us now consider an example in $\mathbb{R}^3$. We first observe
that the parallelepiped can be easily dealt with by extending the
techniques of Sec.\ \ref{sec:strip}. We therefore look at a more
interesting situation. Consider a spherical shell in
$\mathbb{R}^3$ and a domain $D=\{(x,y,z)\;|\;r_1^2\leq x^2+y^2+z^2
\leq r_2^2\}$. The projection on $D$ reads
\barr
\label{eq:projshell}
P=\chi_{[r_1,r_2]}(r)&=&\int_D dx dy dz
\ket{xyz}\bra{xyz} \nonumber \\
&=&\int_{r_1}^{r_2}r^2 dr\int_0^\pi \sin\theta
d\theta\int_0^{2\pi}d\phi \ket{r\theta\phi}\bra{r\theta\phi}.
\earr
The Zeno Hamiltonian, engendering the Zeno subdynamics, is given
by (\ref{eq:HZ}) and represents a free particle in the spherical
shell $D$ with Dirichlet boundary condition
\barr
\label{eq:HZshell}
H_Z&=&-\frac{\hbar^2}{2M}(\partial_x^2+\partial_y^2+\partial_y^2)
\\
& = & -\frac{\hbar^2}{2M r^2}\left(\partial_r(r^2
\partial_r)+\frac{1}{\sin\theta}\partial_\theta(\sin\theta\partial_\theta)
+\frac{1}{\sin^2\theta}\partial_\phi^2\right)
, \nonumber\\
& & \quad \psi(r_1,\theta,\phi)=\psi(r_2,\theta,\phi)=0.
\label{eq:HZshellb}
\earr
As is well known, by writing the eigenfunctions of $H_Z$ as
$\Psi_{nlm}(r,\theta,\phi)=R_{nl}(r)Y_{lm}(\theta)\Phi_m(\phi)$,
the radial part of the eigenvalue equation reads
\barr
\label{eq:eigenshell}
& & -\frac{\hbar^2}{2M}\frac{1}{r^2}\partial_r(r^2
\partial_r)R_{nl}(r)+\frac{\hbar^2}{2M} \frac{l(l+1)}{r^2}R_{nl}(r)=E_{nl}R_{nl}(r),
\label{eq:eigenshellb} \\
& & \qquad R_{nl}(r_1)=R_{nl}(r_2)=0
\earr
and can be solved in terms of spherical Bessel functions.

Let us see how one can obtain $H_Z$ in this case. The first steps
of the derivation are the same as before. By rewriting the
distance $d(r'\theta'\phi',r\theta\phi)$ as
\beq
d^2=(r'-r)^2+2r'r(1-\cos(\theta'-\theta))+2r'r\sin\theta'\sin\theta(1-\cos(\phi'-\phi))
\eeq
it is apparent that the integrals must be performed in the order
$\phi \to \theta \to r$ and that only those stationary points that
do not give an additional phase contribute to the final result.

As eigenfunctions we choose the orthogonal set
\beq
\Psi_{nlm}(r\theta\phi)=R_{nl}(r)Y_{lm}(\theta)\Phi_m(\phi).
\eeq
The transition element is
\barr
G_{n'l'm',nml}(t)&=&\int_{r_1}^{r_2}r'^2dr'\int_{r_1}^{r_2}r^2drR^*_{n'l'}(r')R_{nl}(r)
\nonumber \\
& & \times \sqrt{\frac{M}{2\pi i \hbar
t}}e^{i\frac{M(r'-r)^2}{2\hbar t}}
\frac{1}{r'r}G_{l'm',lm},\\
G_{l'm',lm}&=&\int_0^\pi\sin\theta' d\theta'\int_0^\pi\sin\theta
d\theta Y^*_{l'm'}(\theta')Y_{lm}(\theta)
\nonumber \\
& & \times \sqrt{\frac{Mr'r}{2\pi i \hbar t}}e^{i\frac{M}{\hbar
t}r'r(1-\cos(\theta'-\theta))}\nonumber\\
&&\times\frac{1}{\sqrt{\sin\theta'\sin\theta}}
\int_0^{2\pi}d\phi'\int_0^{2\pi}d\phi\sqrt{\frac{Mr'\sin\theta'r\sin\theta}{2\pi
i \hbar t}}\nonumber\\
&&\times e^{i\frac{M}{\hbar
t}r'r\sin\theta'\sin\theta(1-\cos(\phi'-\phi))}\Phi^*_{m'}(\phi')\Phi_m(\phi)
\earr
The $\phi',\phi$ integral is immediately computed as in the case
of the annulus, Sec.\ \ref{sec:annulus}. $\Phi_m$ must therefore
satisfy the differential equation
\barr
-\Phi_m''=\alpha_m\Phi_m,\qquad \Phi_m(0)=\Phi_m(2\pi),
\earr
so that $\alpha_m=m^2$. Then $G_{l'm'lm}$ becomes
\barr
G_{l'm',lm}&=&\int_0^\pi
\sqrt{\sin\theta'}d\theta'\int_0^\pi\sqrt{\sin\theta} d\theta
Y^*_{l'm'}(\theta')Y_{lm}(\theta)\sqrt{\frac{Mr'r}{2\pi i
\hbar t}}\nonumber\\
&&\times e^{i\frac{M}{\hbar
t}r'r(1-\cos(\theta'-\theta))}\left(1-i\frac{\hbar t}{2M
r'\sin\theta' r\sin\theta}(m^2-1/4)\right)\delta_{m'm}. \nonumber
\\
\earr
The integral over $\theta',\theta$ can be computed in a standard
way (do not forget the $\xi^4$ term in the cosine series) and this
in turn requires that the function
$A_{lm}=\sqrt{\sin\theta}Y_{lm}$ must satisfy the differential
equation
\beq
A''_{lm}+\frac{1}{4}A_{lm}-\frac{m^2-1/4}{\sin^2\theta}A_{lm}=-\alpha_{lm}A_{lm},
\eeq
or, equivalently,
\beq
\frac{1}{\sin\theta}\frac{\partial}{\partial\theta}\left(\sin\theta\frac{\partial}
{\partial\theta}\right)Y_{lm}-\frac{m^2}{\sin^2\theta}Y_{lm}=-\alpha_{lm}Y_{lm},
\eeq
with $Y_{lm}(0)=Y_{lm}(\pi)$. This is the standard equation for
spherical harmonics. It is known that $\alpha_{lm}=l(l+1)$
irrespectively of the value of $m$. We obtain
\barr
G_{n'l'm'nlm}&=&\int_{r_1}^{r_2}r'dr'\int_{r_1}^{r_2}rdr\;R^*_{n'l'}(r')R_{nl}(r)
\sqrt{\frac{M}{2\pi
i \hbar t}}e^{i\frac{M(r'-r)^2}{2\hbar t}}
\nonumber \\
& & \times
\left(1-i\frac{\hbar
l(l+1)}{2Mr'r}t\right)\delta_{l'l}\delta_{m'm}.
\earr
Finally, the differential equation for $A_{nl}=rR_{nl}$ reads
(here $E_{nl}=\frac{\hbar^2 k^2_{nl}}{2M}$, which is independent
of $m$)
\barr
-A''_{nl}+\frac{l(l+1)}{r^2}A_{nl}=k^2_{nl}A_{nl},\qquad
A_{nl}(r_1)=A_{nl}(r_2)=0,
\earr
or, equivalently, in terms of $R_{nl}$, Eq.\
(\ref{eq:eigenshell}). The Zeno Hamiltonian is therefore given by
(\ref{eq:HZshell})-(\ref{eq:HZshellb}).

\section{The general case}
\andy{generalization}
\label{sec:generalization}

By looking at the preceding examples one might think that the
method introduced in this article is parochial and works only, for
example, when the domain, besides being sufficiently regular, is
also endowed with particular symmetries (regular polygons,
circles, spheres and so on), that enable one to introduce
coordinates with a range of integration that can be reduced to a
product of intervals. In turn, this might appear as an implicit
condition of separability, e.g.\ in the case of the
three-dimensional Schr\"odinger equation \cite{Pauling}. On the
contrary, as will be shown in this section, the method we propose
is of general applicability.

Consider again the Hamiltonian (\ref{eq:HU}) and the projection
(\ref{eq:P}), $D\subset\mathbb{R}^N$ being a compact domain with
nonempty interior and a regular boundary. The $N$-dimensional
propagator (\ref{eq:matrixel}) reads
\barr
G_{\bm m ,\bm n}(t)&=& \bra{\bm m}G(t)\ket{\bm n} \nonumber \\
 & = & \int_D d^N x\int_D d^N y
\left(\frac{M}{2\pi i\hbar t}\right)^{N/2}e^{i\frac{M(\bm x-\bm
y)^2}{2\hbar t}} \Psi^*_{\bm m}(\bm x)\Psi_{\bm n}(\bm y)
\label{eq:Ndimprop}
\earr
and by substituting $\bm \xi=\bm y-\bm x$ one gets
\barr
G_{\bm m , \bm n}(t)&=&\int_D d^N x\; \Psi_{\bm m}^*(\bm x)
\int_{D-\bm x} d^N  \xi \left(\frac{M}{2\pi i\hbar
t}\right)^{N/2}e^{i\frac{M \bm \xi^2}{2\hbar t}}
\Psi_{\bm n}(\bm x+\bm \xi) \nonumber \\
&=& \int_D d^N  x \;\Psi^*_{\bm m}(\bm x)
[\mbox{bound}+\mbox{stat}],
\label{eq:nnn}
\earr
where
\beq
D-\bm x=\{\bm y \;|\; \bm x + \bm y \in D \}.
\eeq
Let us evaluate separately the two contributions in the small-$t$
limit. In order to compute the boundary term, we first observe
that
\beq
e^{i \lambda \bm \xi^2} = \frac{\bm \xi \cdot \nabla e^{i \lambda
\bm \xi^2}}{2i\lambda \bm \xi^2}
\label{eq:vecfield}
\eeq
and then integrate by parts ($\lambda=M/2\hbar t$)
\barr
\mbox{bound}&=& \int_D d^N \xi \left(\frac{\lambda}{\pi
i}\right)^{N/2} \Psi_{\bm n}(\bm x+\bm \xi) \frac{\bm \xi \cdot
\nabla e^{i \lambda \bm \xi^2}}{2i\lambda \bm \xi^2}
 \nonumber \\
& = & \left(\frac{\lambda}{\pi i}\right)^{N/2} \left[ \int_D d^N
\xi\;
 \nabla\cdot \left(\frac{\Psi_{\bm n}(\bm x+\bm \xi) \bm \xi e^{i \lambda \bm \xi^2}}
{2i\lambda\bm \xi^2}\right) \right.\nonumber\\
& & \qquad\qquad
 \left. - \int_D d^N \xi\;
 \nabla\cdot \left(\frac{\Psi_{\bm n}(\bm x+\bm \xi) \bm \xi}{\bm \xi^2}\right) \frac{\bm \xi \cdot
\nabla e^{i \lambda \bm \xi^2}}{(2i\lambda)^2 \bm \xi^2} \right]
\nonumber \\
& = & \left(\frac{\lambda}{\pi i}\right)^{N/2} \left[
\oint_{\partial (D -\bm x)} d^{N-1}S\;
 \frac{\Psi_{\bm n}(\bm x+\bm \xi) \bm \xi \cdot \hat{\bm u}}
{\bm \xi^2} \; \frac{e^{i \lambda \bm \xi^2}}{2i\lambda} \left(1+
O(\lambda^{-1})\right) \right]
\nonumber \\
& = & \left(\frac{M}{2\pi i\hbar t}\right)^{N/2}
\left[\oint_{\partial D} d^{N-1}S\; \frac{\Psi_{\bm n}(\bm y) (\bm
y -\bm x) \cdot \hat{\bm u}} {(\bm y -\bm x)^2 } \; \frac{e^{i M
(\bm x -\bm y)^2/2\hbar t}}{iM
/\hbar t }   \right. \nonumber \\
& & \qquad \qquad \qquad \left. \left(1+ \; O(t)\right) \right],
\label{eq:bounbb}
\earr
$\hat{\bm u}$ being the unit vector perpendicular to the boundary.
In the second equality, Eq.\ (\ref{eq:vecfield}) was used again in
order to obtain a higher-order volume integral with the same
structure as the initial one. The stationary contribution is
obtained, as usual, by expanding the integrand function around
$\bm x$
\barr
\mbox{stat}&=& \left(\frac{M}{2\pi i\hbar t}\right)^{N/2}
\int d^N \xi\; e^{i \lambda \bm \xi^2} \nonumber \\
 & & \times \left(\Psi_{\bm n}(\bm  x)+
\nabla \Psi_{\bm n}(\bm x)\cdot \bm \xi+ \frac{1}{2!}
\partial_i\partial_j \Psi_{\bm n}(\bm x)\xi_i \xi_j +
O(|\bm\xi|^3)\right).
\earr
Observe that the contributions of the linear and quadratic (with
$i\neq j$) terms in the integral vanish due to symmetry and one is
left with
\barr
\mbox{stat}&=& \Psi_{\bm n}(\bm x)+i\frac{t\hbar}{2M}\triangle
\Psi_{\bm n}(\bm x)+O(t^{2}).
\earr
In order to obtain (\ref{eq:diag})-(\ref{eq:diagbis}) from
(\ref{eq:nnn})
 one must require that the leading contribution in the boundary
 term (\ref{eq:bounbb}) vanishes and
\barr
-\frac{\hbar}{2M} \triangle\Psi_{\bm n}(\bm x)&=&\frac{E_{\bm
n}}{\hbar}\Psi_{\bm n}(\bm x),
\earr
namely
\barr
\label{eq:diffbcc}
-\frac{\hbar^2 }{2M} \triangle \Psi_{\bm n}(\bm x)=E_{\bm
n}\Psi_{\bm n}(\bm x), \qquad \mathrm{with}\quad \Psi_{\bm
n}(\partial D)=0 .
\earr
Notice that the Schr\"{o}dinger equation is obtained from the
stationary contribution to the asymptotic expansion, while the
Dirichlet boundary conditions are a consequence of the requirement
that the boundary term (\ref{eq:bounbb}) vanish at the lowest
order in the expansion.

Let us briefly comment on the features of the method introduced.
As already emphasized at the end of Section \ref{sec:strip}, this
analysis, although not entirely rigorous, yields the correct
result. We derived the desired properties of the propagator by
requiring at the same time the validity of the Schr\"odinger
equation and the Dirichlet (hard-wall) boundary conditions for the
eigenbasis of the (Zeno) Hamiltonian. We should emphasize,
however, that the boundary and stationary terms are being dealt
with separately. In fact, we did not consider the contribution of
those boundary points that are \emph{also} stationary points. Such
points belong to the intersection of the boundaries of the two
domains $D$ in (\ref{eq:Ndimprop}), namely $\bm x = \bm y \in
\partial D$, and should be analyzed with great care.
A more rigorous treatment can be given, in which the contribution
of the integral (\ref{eq:nnn}) is uniformly estimated: this
analysis requires a different evaluation of the boundary terms and
will be presented elsewhere.

The introduction of a potential \cite{regularize} is not difficult
to deal with if the detailed features of the convergence
(\ref{eq:matrixel})-(\ref{eq:diag}) are not worked out. Much
additional care is required at a deeper mathematical level, when
the self adjointness of the Hamiltonian is called into question
and must be explicitly proved. If additional rigorous results
\cite{MS,Friedman72,Gustavson,ExnerIchinose}
are taken into account and, by an educated guess, extended to the
case of a sufficiently regular potential, one is tempted to assume
that the procedure sketched above is valid in general and the Zeno
dynamics governed by a self-adjoint generator (and a unitary
group). The situation may clearly become more complicated when the
potential is singular and/or the projected spatial region (or its
boundary) lacks the required regularity.

\section{Zeno dynamics in Heisenberg picture}
\andy{algebra}
\label{sec:algebra}

In this section we would like to consider the Zeno dynamics in the
framework of the Heisenberg picture. The following discussion must
be considered preliminary: additional details and a broader
picture will be given in a forthcoming paper. An interesting and
natural question concerns the algebra of observables after the
projection. This is not a simple problem. One can assume that to a
given observable $\mathcal{O}$ before the Zeno projection
procedure there corresponds the observable $P\mathcal{O}P$ in the
projected space:
\beq
\mathcal{O} \Rightarrow P\mathcal{O}P.
\label{eq:OPOP}
\eeq
For example, if one starts in $\mathbb{R}$ and projects over a
finite interval $P=\chi_I(x)$ ($I$ being an interval of
$\mathbb{R}$), the momentum and position operators become
\barr
p \Rightarrow PpP = \left\{ \matrix{ i\partial_x & \mbox{for} \; x
\in I \cr
              0  & \mbox{otherwise}}  \right. \ ,\\
x \Rightarrow PxP = \left\{ \matrix{x & \mbox{for} \; x \in I \cr
              0 & \mbox{otherwise}}\right. \ .
\earr
In this respect it is easy to see that the correspondence
(\ref{eq:OPOP}) is \emph{not} an algebra homomorphism. However, if
we redefine a new associative product in the algebra of operators,
by setting
\beq
A*B \equiv APB ,
\label{eq:star}
\eeq
with this new product the previous correspondence (\ref{eq:OPOP})
becomes an algebra homomorphism \cite{MMSZ}. Notice also that the
new (projected) algebra acquires a unity operator $P$.

Notice that in general the evolution will not be an automorphism
of the new product. However, it will respect the product to order
$O(t/N)$ and induce, in the limit, a Zeno dynamics on the
projected algebra, i.e.\ on the image of the projection. The
evolution will be trivially an automorphism when it commutes with
$P$ and is therefore compatible with the new product without any
approximation. For instance, this would be the case if we take as
Hamiltonian the square of the angular momentum in the case of the
annulus (Sec.\
\ref{sec:annulus}).

In general one has to modify the associative product in such a way
that the ``deviation"  of $U(t/N)$ from being an automorphism is
of order $o(t/N)$, so that in the limit $U_Z(t)$ will be an
automorphism of the new associative product adapted to the
constraint. In other words, the sequence of evolution operators
\begin{equation}\label{eq:evseq}
V_N(t)=G(t/N)^N=(P U(t/N) P)^N ,
\end{equation}
yielding the Zeno limit (\ref{eq:UZ}), is mirrored at the level of
the algebra by the following sequence of deformed associative
products
\begin{equation}\label{eq:prodseq}
A *_N B \equiv A P_N B ,
\end{equation}
where $P_N$ is a smooth positive operator with $0\leq P_N\leq 1$
and $P_N P = P P_N= P$. For any $N$, $P_N$ forms together with
$Q_N=1-P_N$ a positive operator valued measure, yielding a
resolution of the identity, i.e.\ $P_N+Q_N=1$, which approximates
the orthogonal resolution $P+Q=1$, in the sense that
\begin{equation}\label{eq:PNord}
P_N \psi =P \psi +O(1/N) , \qquad \forall \psi\in
L^2(\mathbb{R}^N) .
\end{equation}
For any $N$ the evolution $V_N(t)$ is an automorphism of the
product $*_N$ and in the limit $N\to\infty$ we get the desired
result (\ref{eq:star}).

Observe that, for unbounded operators, (\ref{eq:OPOP}) does not
necessarily yield self-adjoint operators: for example, after the
Zeno procedure, the momentum $p$ would act on functions that
vanish on the boundary of $I$ and would have deficiencies $\langle
1,1 \rangle$, see \cite{compact}. On the other hand the Zeno
Hamiltonian (\ref{eq:HZ}) is self adjoint. However, it would be
arbitrary to require a similar property for every observable in
the algebra. We shall analyze this issue in greater detail in a
future article. In general, the lack of self-adjointness of the
operators representing the ``observables" of the system in the
projected subspace might be related to the incompleteness of the
corresponding classical field \cite{Klauder,compact}.

\section{Projections onto lower dimensional regions: constraints}
\andy{tuttoins}
\label{sec:tuttoins}

In all the situations considered so far, the projected domain
always has the same dimensionality of the original space
($\mathbb{R}^n$). [Remember that, after Eq.\ (\ref{eq:HU}), we
required the projected domain $D$ to have a nonempty interior.]
However, it is interesting to ask what would happen if one would
project onto a domain $D'$ of lower dimensionality. This is
clearly a more delicate problem, as one necessarily has to face
the presence of divergences. It goes without saying that these
divergences must be ascribed to the lower dimensionality of the
projected domain and \emph{not} directly to the convergence
features of the Zeno propagator \cite{Froese}. Our problem is to
understand how these divergences can be cured. One way to tackle
this problem is to start from a projection onto a domain $D
\subset \mathbb{R}^n$ and {\it then} take the limit $D \to
D'\subset \mathbb{R}^{n-1}$, with a Hilbert space (Zeno subspace)
$L^2(D')$.

The content of this section is preliminary. We shall only sketch
the main ideas and postpone a thorough analysis to a forthcoming
paper, in which the physical meaning of the divergences will be
spelled out in greater details.

\subsection{From the rectangle to the interval}
\andy{interval}
\label{sec:interval}

Let us first look at the case of the rectangle, investigated in
Sec.\ \ref{sec:strip}, and let  $b\to 0$. We first notice that in
order to get a sensible result one must \emph{first} perform the
Zeno limit $N\to\infty$ and \emph{then} let $b\to 0$. In
particular one must require
\beq
\delta t=t/N\ll \hbar/E_m=\frac{2M b^2}{\hbar m^2},
\eeq
which has an appealing physical meaning: the time during which the
particle evolves freely between two projections must be small
enough so that the particle remains well within the rectangle of
width $b$. In practice, one must first set $m<m^*$, for some
$m^*$, in order to obtain a sensible result and finally let $m^*$
become arbitrarily large. The order in which the two limits
($N\to\infty$ and $b\to 0$) are taken is relevant and significant
from a physical perspective: one must first make sure that the
wave function does not ``leak" out of the projected (Zeno) region
and then let this region ``shrink" into a domain of lower
dimensionality.

However, even if one follows the correct procedure (i.e., first
$N\to\infty$ and then $b \to 0$) one still gets divergences in the
phases, since
\beq
E_{m}=\frac{\hbar^2 \pi^2 m^2}{2M b^2}\to\infty \qquad
\mathrm{for}\quad b\to 0.
\eeq
Notice also that since the energy differences between different
$m$ states diverge, a superselection rule arises. Different
subspaces, labeled by different values of the quantum number $m$,
remain separated (at least for low-energy processes with energies
$E\ll \hbar^2/M b^2$). This is also physically revealing.

On the basis of the above insights, we therefore propose to
perform the limit $b\to 0$ by choosing a particular eigenstate
$\phi_m(y)$ and considering the reduced evolution
\beq
\widetilde U_Z(t)=e^{iE_{m}t/\hbar}\bra{m}U_Z(t)\ket{m},
\eeq
which operates only on the $x$ degree of freedom. Physically, this
corresponds to the choice of a particular value of the
superselection charge. Thus, the reduced propagator reads
\barr
G(x',x;t)=\bra{x'}\widetilde U_Z(t)\ket{x}&=&e^{iE_m
t/\hbar}\bra{m; x'}U_Z(t)\ket{m;x}
\nonumber \\
&=&\sum_n e^{-iE_nt/\hbar}\psi_n(x')\psi_n^*(x),
\earr
where $\psi_n$ are the eigenfunctions of the Dirichlet problem
(\ref{eq:diffbc}), and one gets
\beq
\widetilde H_Z=-\frac{\hbar^2\partial_x^2}{2 M}, \qquad \mbox{with
Dirichlet b.c.}
\eeq
This is just the free particle on the interval $[0,a]$, as
expected. Not only can the divergence be cured, it also yields the
desired result.

The framework explained in this particular example works also in
more complicated circumstances. In particular, it is important to
understand in which order the two limits must be computed: first
one makes sure that the Zeno mechanism works efficaciously, then
takes the desired limit on the domain. We consider here two other
simple situations.

\subsection{From the annulus to the circle}
\andy{circle}
\label{sec:circle}

Let us now look at the annulus, investigated in Sec.\
\ref{sec:annulus}. We would like to recover the evolution of a
particle on a circle by considering the $\delta r\to 0$ limit,
while keeping $r_1=r_2=R$ constant. Once again, as in Sec.\
\ref{sec:interval}, we have to face some divergences. By taking
the limit one finds the approximate eigenfunctions of Eq.\
(\ref{eq:eigenring})
\beq
\psi_{nl}(r)\simeq\psi_n(r)=\sqrt{\frac{2}{R\delta
r}}\sin\left(\frac{n\pi}{\delta r}(r-R)\right)
\eeq
and the energies
\beq
E_{nl}\simeq E_n+E_l=\frac{\hbar^2}{2M}\frac{n^2\pi^2}{\delta
r^2}+  \frac{\hbar^2}{2M R^2}\left(l^2-\frac{1}{4}\right).
\eeq
Again one finds a diverging energy which must be regularized.
However a second (finite) term appears ($-\hbar^2/8 M R^2$)
\cite{Goldstone} which is not present in the usual circle
quantization. We notice that different quantization methods yield
different results \cite{Anto}.

The reduced propagator on the remaining degree of freedom $\theta$
is just
\beq
G(\theta',\theta;t)=e^{i\frac{\hbar^2}{2M}\frac{n^2\pi^2}{\delta
r^2}}\bra{n,\theta'}U_Z(t)\ket{n,\theta}=\sum_l e^{-iE_l
t/\hbar}\phi^*_l(\theta')\phi_l(\theta),
\eeq
which is what one expected.

\subsection{From the shell to the sphere}
Finally, we reconsider the spherical shell of Sec.\
\ref{sec:sphere} and take the limit $\delta r\to 0$, while keeping
$r_1=r_2=R$ constant, like in Sec.\ \ref{sec:circle}. This yields
the energies
\beq
E_{nl}\simeq\frac{\hbar^2 n^2\pi^2}{2m\delta
r^2}+\frac{\hbar^2l(l+1)}{2MR^2}
\eeq
and following the same regularization procedure as before we find
\barr
G(\theta',\phi',\theta,\phi;t)&=&e^{i\frac{\hbar \pi^2
n^2}{2M\delta r^2}t}\bra{n;\theta',\phi'}U_Z(t)\ket{n;\theta,\phi}
\nonumber \\
&=&\sum_{lm}e^{-i\frac{\hbar
l(l+1)}{2MR^2}t}Y_{lm}(\theta')\Phi_m(\phi')Y^*_{lm}(\theta)\Phi^*_m(\phi),
\earr
which is the usual propagator on a sphere of radius $R$, whose
Hamiltonian is
\beq
\tilde H_Z=\frac{\bm L^2}{2 M R^2}=-\frac{\hbar^2}{2M R^2}
\left(\frac{1}{\sin\theta}\partial_\theta(\sin\theta\partial_\theta)
+\frac{1}{\sin^2\theta}\partial_\phi^2\right).
\eeq

\section{Concluding remarks on potential applications}
\andy{concl}
\label{sec:concl}

We have investigated the quantum Zeno dynamics, when a free system
undergoes frequent measurements that ascertain whether it is
within a sufficiently regular spatial region. The evolution in the
projected (Zeno) subspace is unitary and the generator of the Zeno
dynamics is the Hamiltonian with hard-wall (Dirichlet) boundary
conditions. In general, this procedure leads to the formation of
the ``Zeno subspaces" \cite{FPsuper}, on whose boundaries the wave
function must vanish (Dirichlet): this is the ultimate reason for
the absence of amplitude (and probability) leakage between
``adjacent" subspaces.

Quantum computation \cite{Review} is one of the most promising
fields of potential application of the QZE. Interactions with the
environment deteriorate the purity of quantum states and represent
a very serious obstacle against the preservation of quantum
superpositions and entanglement over long periods of time. It is
therefore of great interest to endeavor to understand whether
decoherence can be controlled and eventually halted
\cite{deccontrol}: in this context, novel techniques hinging upon
the quantum Zeno effect are of interest. Besides the use of
quantum error correcting codes \cite{ErrorCorrecting}, the
engineering of ``decoherence-free" subspaces is also recently
being considered and widely investigated \cite{NoiselessSub}. Some
mechanisms are actually being proposed, based on the so-called
``bang-bang" evolutions and their generalization, quantum
dynamical decoupling \cite{BBDD}. Although ``bang-bang" techniques
in \emph{classical} control theory are know to engineers since
long ago \cite{BBclass}, their introduction as a quantum control
and their unification with the basic ideas underlying the quantum
Zeno effect are quite recent \cite{bang}. In particular, the
decoherence-free subspaces are the dynamically generated quantum
Zeno subspaces \cite{FPsuper} within which the dynamics is far
from being trivial, as has been discussed in this article. It is
also worth noticing that the range of applicability of these ideas
is wide, as frequent interruptions and continuous coupling
\cite{cont} can yield similar physical effects. This is not
entirely surprising \cite{QZEbxl}, in view of Wigner's notion of
``spectral decomposition" \cite{Wigner63}. However, when one
considers applications of the Zeno dynamics in the context of
decoherence-free subspaces, one must remember that if the
measurement is not \emph{very} frequent, the quantum evolution
yields the so-called ``inverse" or ``anti" Zeno effect, by which
transitions out of the decoherence-free subspace is accelerated
\cite{antiZeno}.

In conclusion, it is interesting to notice that an issue that was
considered as purely academic until a few years ago, has been
first experimentally demonstrated and is now being considered as a
possible strategy to combat decoherence, with interesting spinoffs
and very practical applications.


\section*{References}

\begin{thebibliography}{99}

\bibitem{QZEseminal}
von Neumann J 1955 {\it Mathematical Foundation of Quantum
Mechanics} (Princeton: Princeton University Press) \nonum Beskow A
and Nilsson J 1967 {\it Arkiv f\"ur Fysik} {\bf 34} 561 \nonum
Khalfin L A 1968 {\it JETP Letters} {\bf 8} 65

\bibitem{MS}
Misra B and Sudarshan E C G 1977 {\it J. Math. Phys.} {\bf 18} 756

\bibitem{QZEreview}
Home D and Whitaker M A B 1997 {\it Ann. Phys.} {\bf 258} 237
\nonum Facchi P and Pascazio S 2001 {\it Progress in Optics} Wolf
E (ed) (Amsterdam: Elsevier) {\bf 42} 147

\bibitem{Schwinger}
Schwinger J 1959 {\it Proc. Natl. Acad. Sci. U.S.} {\bf 45} 1552
\nonum \dash 1970 {\it Quantum Kinetics and Dynamics} (New York:
Benjamin)

\bibitem{compact}
\andy{compact} Facchi P, Gorini V, Marmo G, Pascazio S and
Sudarshan E C G 2000 {\it Phys. Lett.} A \textbf{275} 12

\bibitem{regularize}
\andy{regularize} Facchi P, Pascazio S, Scardicchio A and Schulman
L S 2002 {\it Phys. Rev.} A \textbf{65} 012108

\bibitem{FPsuper}
Facchi P and Pascazio S 2002 {\it Phys. Rev. Lett.} {\bf 89}
080401 \nonum \dash ``Quantum Zeno subspaces and dynamical
superselection rules" 2003 in \textit{Proceedings of the 22nd
Solvay Conference} (Singapore: World Scientific) 251

\bibitem{antiZeno}
Lane A M 1983 {\it Phys. Lett.} A {\bf 99} 359 \nonum Schieve W C,
Horwitz L P and Levitan J 1989 {\it Phys. Lett.} A {\bf 136} 264
\nonum Kaulakys B and Gontis V 1997 \textit{Phys. Rev.} A
\textbf{56} 1131 \nonum Thun K and Pe\v{r}ina J 1998 \textit{Phys.
Lett.} A \textbf{249} 363 \nonum Facchi P and Pascazio S 2000
\textit{Phys. Rev.} A \textbf{62} 023804 \nonum Elattari B and
Gurvitz S A 2000 {\it Phys. Rev.} A {\bf 62} 032102 \nonum Kofman
A G and Kurizki G 2000 {\it Nature} {\bf 405} 546 \nonum Facchi P,
Nakazato H and Pascazio S 2001 {\it Phys. Rev. Lett.} {\bf 86}
2699 \nonum Koshino K and Shimizu A 2003 {\it Phys. Rev.} A {\bf
67} 042101

\bibitem{WinelandZeno}
Itano W M,  Heinzen D J,  Bolinger J J and Wineland D J 1990 {\it
Phys. Rev.} A {\bf 41} 2295

\bibitem{Wilkinson} \andy{Wilkinson}
Wilkinson S R, Bharucha C F,  Fischer M C, Madison K W,  Morrow P
R, Niu Q, Sundaram B and  Raizen M G 1997 \textit{Nature}
\textbf{387} 575

\bibitem{raizenlatest} Fischer M C,  Guti\'errez-Medina B and
Raizen M G 2001 \textit{Phys. Rev. Lett.} \textbf{87} 040402

\bibitem{Balzer} \andy{Balzer}
Balzer C,  Huesmann R, Neuhauser W and Toschek P E 2000
\textit{Opt. Comm.} \textbf{180} 115

\bibitem{valanju} Valanju P, Sudarshan E C G and Chiu C B 1980
{\it Phys. Rev.} D {\bf 21} 1304

\bibitem{Cook}
Cook R J 1988 {\it Phys. Scr.} T {\bf 21} 49

\bibitem{Itanodisc} \andy{Itanodisc}
Peres A and Ron A 1990 {\it Phys. Rev.} A {\bf 42} 5720 \nonum
Itano W H, Heinzen D J, Bollinger J J and Wineland D J 1991 {\it
Phys.\ Rev.} A {\bf 43} 5168 \nonum Inagaki S, Namiki M and Tajiri
T 1992 {\it Phys. Lett.} A {\bf 166} 5 \nonum Pascazio S, Namiki
M, Badurek G and Rauch H 1993 {\it Phys.\ Lett.} A {\bf 179} 155
\nonum Blanchard Ph and Jadczyk A 1993 {\it Phys.\ Lett.} A {\bf
183} 272 \nonum Altenm\"uller T P and Schenzle A 1994 {\it Phys.\
Rev.} A {\bf 49} 2016 \nonum Berry M 1995 {\it Fundamental
Problems in Quantum Theory} Greenberger D M and Zeilinger A (ed)
(New York: Ann. N.Y. Acad. Sci.) {\bf 755} 303 \nonum Beige A and
Hegerfeldt G 1996 {\it Phys. Rev.} A {\bf 53} 53 \nonum Luis A and
Pe\v rina J 1996 \textit{Phys. Rev. Lett.} \textbf{76} 4340

\bibitem{Friedman72}
\andy{Friedman72}
Friedman C N 1972 {\it Indiana Univ. Math. J.} \textbf{21} 1001

\bibitem{Gustavson}
\andy{Gustavson} Gustafson K 1983 ``Irreversibility questions in
chemistry, quantum-counting, and time-delay" in \textit{Energy
storage and redistribution in molecules}, Hinze J (ed) (Plenum)
and refs. [10,12] therein

\bibitem{MisraAntoniou}
\andy{MisraAntoniou} Misra B and Antoniou A 2003 ``Quantum Zeno
effect" in \textit{Proceedings of the 22nd Solvay Conference}
(Singapore: World Scientific) 233

\bibitem{GustavsonSolvay}
\andy{GustavsonSolvay}
Gustafson K 2003 ``A Zeno story" {\it Preprint} quant-ph/0203032

\bibitem{Schmidt}
\andy{Schmidt} Schmidt A U 2002 {\it J. Phys.} A: {\it Math. Gen.}
{\bf 35} 7817 \nonum \dash 2003 {\it J. Phys.} A: {\it Math. Gen.}
{\bf 36} 1135

\bibitem{ExnerIchinose}
\andy{ExnerIchinose}
Exner P and Ichinose T 2003 ``Product formula related to quantum
Zeno dynamics" {\it Preprint} math-ph/0302060

\bibitem{Sommerfeld}
Sommerfeld A 1949 {\it Partial differential equation in physics}
(New York: Academic Press) p 168

\bibitem{Pauling}
\andy{Pauling} Pauling L and Wilson E B 1935 {\it Introduction to
quantum mechanics} (Singapore: McGraw-Hill) 443 \nonum  Eisenhart
L P 1934 {\it Phys. Rev.} {\bf 45} 428


\bibitem{MMSZ}
\andy{MMSZ} Man'ko V I, Marmo G, Zaccaria F and Sudarshan E C G
1997 \textit{Int. J. Mod. Phys.} B \textbf{11} 1281 \nonum
Carinena J F, Grabowski J and Marmo G 2000 \textit{Int. J. Mod.
Phys.} A \textbf{15} 4797

\bibitem{Klauder}
\andy{Klauder} Nelson E 1969 \textit{Topics in Dynamics I: Flows},
Mathematica Notes (Princeton: Princeton University Press) p 114
\nonum Zhu C and Klauder J R 1993 {\it Am. J. Phys.} {\bf 61} 605

\bibitem{Froese}
\andy{Froese}
Froese R and Herbst I 2001 \textit{Commun. Math. Phys.}
\textbf{220} 489 and references therein

\bibitem{Goldstone}
\andy{Goldstone} Goldstone J and Jaffe R L 1992 \textit{Phys.
Rev.} B \textbf{45} 14100

\bibitem{Anto}
\andy{Anto} Scardicchio A 2002 {\it Phys. Lett.} A {\bf 300} 7

\bibitem{Review} Galindo A and
Martin-Delgado M A 2002 {\it Rev. Mod. Phys.} {\bf 74} 347 \nonum
Bouwmeester D, Ekert A and Zeilinger A (ed) 2000 {\it The Physics
of Quantum Information} (Berlin: Springer) \nonum Nielsen M A and
Chuang I L 2000 {\it Quantum Computation and Quantum Information}
(Cambridge: Cambridge University Press)

\bibitem{deccontrol}
\andy{deccontrol} Kofman A G and Kurizki G 2002 {\it Phys. Rev.
Lett.} \textbf{87} 270405 \nonum Calarco T, Datta A, Fedichev P,
Pazy E and Zoller P 2003 {\it Phys. Rev.} A \textbf{68} 012310
\nonum Tasaki S, Tokuse A, Facchi P and Pascazio S 2004 {\it Int.
J. Quant. Chem.} in print, {\it Preprint} quant-ph/0210129

\bibitem{ErrorCorrecting} Shor P W 1995 {\it Phys. Rev.} A
{\bf 52}, 2493 \nonum Calderbank A R and Shor P W 1996 {\it Phys.
Rev.} A {\bf 54} 1098 \nonum Steane A 1996 {\it Proc. R. Soc.
London} A {\bf 452} 2551 \nonum \dash 1996 {\it Phys. Rev. Lett.}
{\bf 77} 793 \nonum For a review, see Preskill J 1999 in
\textit{Introduction to Quantum Computation and Information} Lo H
K,  Popescu S and Spiller T P (ed) (Singapore: World Scientific)

\bibitem{NoiselessSub}
Palma G M, Suominen K A and Ekert A K 1996 {\it Proc. R. Soc.
Lond.} A \textbf{452} 567 \nonum Duan L M and Guo G C 1997 {\it
Phys. Rev. Lett.} {\bf 79} 1953 \nonum Zanardi P and Rasetti M
1997 {\it Phys. Rev. Lett.} {\bf 79} 3306 \nonum Lidar D A, Chuang
I L and Whaley K B 1998 {\it Phys. Rev. Lett.} {\bf 81} 2594
\nonum Knill E, Laflamme R and Viola L 2000 {\it Phys. Rev. Lett.}
{\bf 84} 2525 \nonum For a review, see Lidar D A and Whaley K B
2003 ``Decoherence-Free Subspaces and Subsystems" {\it Preprint}
quant-ph/0301032

\bibitem{BBDD}
Viola L and Lloyd S 1998 \textit{Phys. Rev.} A \textbf{58} 2733
\nonum Viola L,  Knill E and Lloyd S 1999 \textit{Phys. Rev.
Lett.} {\bf 82} 2417 \nonum \dash 1999 \textit{Phys. Rev. Lett.}
{\bf 83} 4888 \nonum \dash 2000 \textit{Phys. Rev. Lett.} {\bf 85}
3520 \nonum Zanardi P 1999 \textit{Phys. Lett.} A {\bf 258} 77
\nonum Vitali D and Tombesi P 1999 \textit{Phys. Rev.} A {\bf 59}
4178 \nonum \dash 2001 \textit{Phys. Rev.} A {\bf 65} 012305
\nonum Uchiyama C and Aihara M 2002 \textit{Phys. Rev.} A
\textbf{66} 032313 \nonum Byrd M S and Lidar D A 2002
\textit{Quantum Information Processing} {\bf 1} 19 \nonum \dash
2003 \textit{Phys. Rev.} A {\bf 67} 012324

\bibitem{BBclass}
Jurdjevic V 1997 \textit{Geometric control theory} (Cambridge:
Cambridge University Press)

\bibitem{bang}
Facchi P, Lidar D A  and Pascazio S 2004 \textit{Phys. Rev.} A
\textbf{69} 0323XX

\bibitem{cont} \andy{cont}
Simonius M 1978 \textit{Phys. Rev. Lett.} \textbf{40} 980 \nonum
Peres A 1980 \textit{Am. J. Phys.} {\bf 48} 931 \nonum Harris R A
and Stodolsky L 1982 \textit{Phys. Lett.} B \textbf{116} 464
\nonum Venugopalan A and Ghosh R 1995 \textit{Phys. Lett.} A
\textbf{204} 11 \nonum Plenio M P, Knight P L and Thompson R C
1996 \textit{Opt. Comm.} \textbf{123} 278 \nonum Berry M V and
Klein S 1996 \textit{J. Mod. Opt.} \textbf{43} 165 \nonum Mihokova
E, Pascazio S and Schulman L S 1997 \textit{Phys. Rev.} A
\textbf{56} 25 \nonum Luis A and S\'{a}nchez--Soto L L 1998
\textit{Phys. Rev.} A \textbf{57} 781 \nonum Schulman L S 1998
\textit{Phys. Rev.} A \textbf{57} 1509 \nonum Thun K and
Pe\v{r}ina J 1998 \textit{Phys. Lett.} A \textbf{249} 363 \nonum
Panov A D 1999 \textit{Phys. Lett.} A  \textbf{260} 441 \nonum
\v{R}eh{\'a}\v{c}ek J, Pe\v{r}ina J, Facchi P, Pascazio S and
Mi\v{s}ta L 2000 \textit{Phys. Rev.} A \textbf{62} 013804 \nonum
Facchi P and Pascazio S 2000 \textit{Phys. Rev.} A \textbf{62}
023804 \nonum Militello B, Messina A and Napoli A 2001
\textit{Phys. Lett.} A \textbf{286} 369 \nonum Luis A 2001
\textit{Phys. Rev.} A \textbf{64} 032104

\bibitem{QZEbxl}
Petrosky T, Tasaki S and Prigogine I 1990 \textit{Phys. Lett.} A
{\bf 151} 109 \nonum \dash  1991 \textit{Physica} A {\bf 170} 306
\nonum Pascazio S and Namiki M 1994 \textit{Phys. Rev.} A
\textbf{50} 4582

\bibitem{Wigner63} \andy{Wigner63}
Wigner E P 1963 \textit{Am. J. Phys.} \textbf{31} 6

\end{thebibliography}
\end{document}